\documentclass[10pt,conference,letterpaper]{IEEEtran} 

\usepackage{cite}
\usepackage{amsmath,amssymb,amsfonts}
\usepackage{algorithm}
\usepackage{algorithmic}
\usepackage{graphicx}
\usepackage{textcomp}

\usepackage{lscape, latexsym}
\usepackage{graphicx}
\usepackage{epstopdf}
\usepackage{caption}
\usepackage{subcaption}
\usepackage{float}

\def\BibTeX{{\rm B\kern-.05em{\sc i\kern-.025em b}\kern-.08em
    T\kern-.1667em\lower.7ex\hbox{E}\kern-.125emX}}

\graphicspath{{figure/}}

\begin{document}

\title{Size Matters: A Comparative Analysis of Community Detection Algorithms\\
}

\author{\IEEEauthorblockN{Paul Wagenseller III }
\IEEEauthorblockA{\textit{School of Mathematical and Natural Sciences} \\
\textit{Arizona State University}\\
Paul.Wagenseller@asu.edu}
\and
\IEEEauthorblockN{Feng Wang}
\IEEEauthorblockA{\textit{School of Mathematical and Natural Sciences} \\
\textit{Arizona State University}\\
fwang25@asu.edu}}



\maketitle
\thispagestyle{empty}

\begin{abstract}

Understanding community structure of social media is critical due to its broad applications such as friend recommendations, user modeling and content personalizations. Existing research uses structural metrics such as modularity and conductance and functional metrics such as ground truth to measure the qualify of the communities discovered by various community detection algorithms, while overlooking a natural and important dimension, \emph{community size}. Recently, anthropologist Dunbar suggests that the size of stable community in social media should be limited to 150, referred to as Dunbar's number. In this study, we propose a systematic way of algorithm comparison by orthogonally integrating community size as a new dimension into existing structural metrics for consistently and holistically evaluating the community quality in social media context. we design a heuristic clique based algorithm which controls the size and overlap of communities with adjustable parameters and evaluate it along with five state-of-the-art community detection algorithms on both Twitter network and DBLP network.
Specifically, we divide the discovered communities based on their size into four classes called close friend, casual friend, acquaintance, and just\textunderscore a\textunderscore face, and then calculate the coverage, modularity, triangle participation ratio, conductance, transitivity, and the internal density of communities in each class. We discover that communities in different classes exhibit diverse structural qualities and many existing community detection algorithms tend to output extremely large communities.  
\end{abstract}


\section{Introduction}
Community is a natural and fundamental element that exists in a wide variety of networked systems, 
such as biology networks, social networks, and neural networks. Identifying communities in these networks is a crucial step for gaining an in-depth understanding on network structure, dynamics and interactions. However, there is no unique and widely accepted goodness measure of community quality in literature. Informally, a good community is a densely-connected group of nodes that is sparsely connected to the rest of the network. Extensive research has been devoted to designing community detection algorithms to uncover communities with the goal to minimize or maximize structural metrics, such as modularity, triangle participation ratio, or conductance of the discovered communities. Many comparative research~\cite{Santo2010,WangVLDB2015, YangNature2016, Harenberg2014, Jure2010} has also been conducted to evaluate and compare state-of-the-art community detection algorithms from the perspective of different structural metrics. In parallel, ~\cite{Yang2012} validates the communities discovered by different community detection algorithms with a functional metric called ground truth, which is the a priori knowledge of user's community membership.


However, the majority of previous research in community detection algorithm design and evaluation has  overlooked an important metric, \emph{community size}, which has significant meaning in the context of social network. The primary objective of forming community in social network is for social interaction and influence, while it is hard for large communities to achieve such goal. Anthropologist Dunbar~\cite{Dunbar2016} suggests that the size of communities with strong ties in both traditional social networks and Internet-based social networks should be limited to 150 (called Dunbar's number) due to the cognitive constraint and time constraint of human being. Large communities of size over 150 contain weak connections among their members therefore are not stable, while small communities of size 2 or 3 cannot provide the strong sense of team or community. Therefore, we refer to communities of size 4 to 150 as \emph{desirable community} in this paper and carry out extensive experiments to systematically evaluate and compare various community detection algorithms taking the size of identified communities into consideration. 

To make the comparison results reproducible, we adopt the following five well known community detection algorithms implemented in igraph package: Infomap~\cite{Infomap2008}, Multilevel~\cite{MultilevelBlonde2008}, FastGreedy~\cite{FastGreedy2004}, Eigenvector~\cite{EigenVector2006}, and Label Propagation~\cite{LabelPropgatoin2007}. Additionally, we propose a clique based algorithm called Clique Augmentation Algorithm (CAA) which augments the cliques in the network into communities while using growing threshold and overlapping threshold to control the size of the community and the amount of overlap among communities. We then evaluate these six algorithms on two datasets with different characteristics, one is a Twitter follower graph and the other is a DBLP network~\cite{snap_dblp}. Specifically, we first study the size distribution of the communities identified by each algorithm, then calculate the percentage of users that are assigned to desirable communities for each algorithm.  Next, we divide the communities into four different size groups and compare the goodness of communities in each group with the following metrics:  1) extended modularity, 2) triangle participation ratio (TPR), 3) conductance, 4) internal density, and 5) transitivity. 

Our experimental results reveal that communities in different groups exhibit diverse structural qualities. We discover that Infomap and CAA outperform others in terms of both community size and coverage by consistently producing desirable communities covering a large portion of users in both Twitter and DBLP networks. Nevertheless, both modularity maximization algorithms including Multilevel, FastGreedy, and Eigenvector and node labeling algorithm Label Propagation tend to output extremely large communities of size over 30,000. Specifically, the largest community identified by Eigenvector holds $71.6\%$ of all users in Twitter graph and $99.5\%$ of all users in DBLP network respectively, therefore lacking practical significance. 

For modularity, Infomap outperforms others because unlike Multilevel, FastGreedy, and Eigenvector, whose modularity scores are mostly decided by the large communities of size [501+], Infomap achieves high score that is contributed by all four groups of it's communities. For conductance, modularity maximization algorithms perform better than other algorithm indicating their identified communities are better separated from the external compared to other algorithms. With regard to triangle participation ratio, CAA outperforms others reaching above $90\%$ in DBLP network and close to $90\%$ in Twitter network. All algorithms perform comparably in terms of internal connectivity metrics, transitivity and internal density, with CAA performs slightly better for desirable community sizes. 

Our contributions in this paper are twofold: 
\begin{enumerate}
\item We introduce Dunbar's number into community detection and evaluation research 
\item We carry out extensive experiment to systematically evaluate and compare various community detection algorithms taking the size of identified communities into consideration. To the best of our knowledge, this is the first effort to investigate this overlooked but important metric in a systematic way.
\end{enumerate}

The remainder of this paper is organized as follows. Section 2 describes the importance of size in social networks. Section 3 introduces the proposed clique augmentation algorithm. Performance comparison is given in Section 4. Section 5 summarizes related work in community detection algorithm design and evaluation, and Section 6 concludes this paper and outlines our future work.

\section{Community Size Matters in Social Media} 
Community size is an important factor in social media. On one hand, groups of size 1, 2, and 3 are too small to be called a community. On the other hand, large communities cannot facilitate communication or interaction therefore members in the community have limited influence upon each other. Dunbar's number of 150, which is the limit of stable online community size, is recently introduced to social media~\cite{Dunbar2016}. It is calculated as a direct function of relative neocortex size which decides the neocortical processing capacity that limits the number of individuals with whom a stable inter-personal relationship can be maintained, and this in turn limits group size. 

Even many online communities have hundreds and even thousands of members, most of those members are inactive, non-participatory in the group on a regular basis. Evidence of Dunbar's number exists ubiquitously, from the size of villages, to the size of modern hunter-gatherer societies, to the number of active members in online gaming communities and online forum, and to the number of active administrators in Wikipedia. Additionally, technical forum tends to break down when it reaches about 80 active contributors, requiring a forum split before continued growth could occur~\cite{Newyorker}.  

By scrutinizing the top 5000 highest quality communities provided in DBLP, a co-authorship network~\cite{snap_dblp}, we found that 4951 of them are of size 4 to 150. This means $99\%$ of the highest quality ground-truth communities in DBLP network conform to Dunbar's number. Furthermore, The dataset claims 13,477 ground truth communities in total and $94\%$ of them have size of 4 to 150. This is another strong evidence of Dunbar's number.  

In addition to the well-known number of 150, ~\cite{Dunbar2016} also gives a series of numbers which correspond to the closeness of a person to people he interacts. According to Dunbar's model, each person can maintain about five people in their support group of closest friends, about 15 people in a sympathy group who are close enough to confide in, about 50 close friends, about 150 casual friends, and about 500 acquaintances. In all, it indicates that any given human can identify about 1,500 faces in total. We illustrate these social circles in Table~\ref{tab:dunbar}. 

\begin{table}[h]
\centering
\caption{Social Circle}
\label{tab:dunbar}\begin{tabular}{|c|c|c|} \hline
   Size  & Closeness  \\ \hline
    4-5 & Support clique \\  \hline
    6-15 & Sympathy group \\ \hline
    16-50 & Close friend \\ \hline
    51-150 & Casual friend \\  \hline
    151-500 & Acquaintance \\ \hline
    501-1500 & Just a face \\ \hline
\end{tabular}
\end{table}

In this paper, seeking to gain a better understanding of how community goodness metrics change with relation to community size, we use these numbers as guideline to divide the identified communities into different size groups.  Specifically, to conform to Dunbar's breakdown and to avoid overcrowded figures for clear presentation, we group the first three categories in Table~\ref{tab:dunbar} into one group. In this way, the six categories are turned into four classes of communities of size [4-50],  [51-150], [151-500], and [501+] respectively. We call these four classes close friend, casual friend, acquaintance, and just\textunderscore a\textunderscore face communities. 

It is worth noting that we are not stating that communities of large size are not important. Large communities are not practical in social media, but they have their own applications in other types of networks such as biological network and neural network. Furthermore, we are not reducing the significance of a community only to its size, but it is worth to give in-depth analysis of the quality of communities an algorithm produces by adding community size as an orthogonal dimension to the existing structural measurements. 

\section{Clique Augmentation Algorithm}
In this section we propose a clique based community detection algorithm called Clique Augmentation Algorithm (CAA). CAA is built on the following two principles: 1) Users in a maximal clique belong to a stable community since a clique is densely connected internally;  2) A neighboring node that is highly connected to a clique should be part of the community since it keeps the triadic closure property among all nodes in the community. 


Given a social network topology, CAA algorithm discovers communities in the topology using the following steps: \\
\indent 1) Find all maximal cliques in the topology; \\
\indent 2) Filter the overlapping cliques.  We sort the cliques based on their size then use an {\em overlapping threshold} to control the amount of overlap between two cliques. The overlapping threshold is defined as the percentage of overlapping nodes in the smaller clique. For example, given two cliques $c_{1}$ and $c_{2}$ where $c_{1}$ is of size 10, and $c_{2}$ is of size 5. Suppose the overlapping threshold is 0.7. If $c_{2}$ only has 2 nodes overlapping with $c_{1}$, we consider $c_{1}$ and $c_{2}$ as two independent cliques since $2 < 5*0.7$ which is 3.5. If $c_{2}$ had 4 overlapping nodes with $c_{1}$, we would discard $c_{2}$ since $4 > 3.5$; \\
\indent 3) Grow each clique into a community by adding new nodes in one by one. {\em Growing threshold} is utilized for controlling the growth of each community. The growing threshold is defined as the ratio of the number of incoming edges from the new node to other nodes in the community over the size of a community. For example, if a community has size of 10, and the growing threshold is set to 0.7, then for a neighboring node to be added into the community, it must have at least 7 edges coming into the community. The algorithm checks the neighboring nodes for each node within the current community. This process is repeated for the updated community until no more nodes can be added. The growing threshold allows us to zoom in or out of the graph around the clique. An analysis of the effect of overlapping and growing threshold is conducted in Section \ref{ssec:growOverlap}. The implementation detail of the algorithm is presented in Algorithm~\ref{CAA}

The time complexity of CAA is dominated by the time complexity of maximal clique algorithm, which is NP-Complete. However, consider the moderate clique size in social network, maximal clique algorithm implemented in NetworkX package runs in polynomial time in practice. After the maximal cliques are obtained, the time to filter overlapping cliques is $O(N^2)$ where $N$ is the number of nodes. The time to augment clique into communities also takes $O(N^2)$. 


\begin{algorithm}[h]
\caption{Clique Augmentation Algorithm}
\label{CAA}
\begin{algorithmic}[1]
\STATE INPUT: A graph $G$, overlapping threshold $\omega$, growing threshold $\phi$
\STATE OUTPUT: communityList in graph $G$ 
\STATE cliques = findMaximalCliques($G$) 
\STATE filteredCliqueList = filterOverlappingCliques($cliques$, $\omega$) 
\STATE foreach clique c in filteredCliqueList:
\STATE \ \  curCommunity = []
\STATE \ \  curCommunity.add(c.nodes())
\STATE \ \  foreach node v in curCommunity:
\STATE \ \ \ \ growingThreshCount = (len(curCommunity) -1) * $\phi$)
\STATE \ \ \ \ neighborList = [] 
\STATE \ \ \ \ foreach neighbor w of v: 
\STATE \ \ \ \ \ \ if w not in curCommunity \&\& w not in neighborList
\STATE \ \ \ \ \ \ \ \ add w to neighborList
\STATE \ \ \ \ growingList = []
\STATE \ \ \ \ foreach node u in neighborList:
\STATE \ \ \ \ \ \ incomingEdgeCount = 0 
\STATE \ \ \ \ \ \ foreach neighbor w of u:
\STATE \ \ \ \ \ \ \ \ if w is in curCommunity  incomingEdgeCount++
\STATE \ \ \ \ \ \ if incomingEdgeCount $\ge$ growingThreshCount
\STATE \ \ \ \ \ \ \ \ add node u to growingList
\STATE \ \ \ \ curCommunity.add(growingList)
\STATE \ \  communityList.add(curCommunity)
\end{algorithmic}
\end{algorithm}

It is worth noting that CAA takes a different approach than Clique Percolation Method (CPM) where two adjacent cliques are merged into a community structure. In CAA, instead of merging neighboring cliques, we simply grow the community structure by adding individual node to the community sequentially. CAA has a few nice features: 1) it is faster than CPM and manages to produce similar results. 2) CAA captures the natural growth process of a community in the sense that if a user befriends with many users in a densely connected community, the user will be most likely grow as part of the community.

\section{Empirical Analysis}

\subsection{Description of Data Sets}
We carry out the comparative research on two datasets: 1) Twitter user follower topology collected over a 3-month period in summer 2013. It contains $318,233$ Twitter users with $3,545,258$ directed edges. Because not every community detection algorithm we measure in this paper supports directed graph, we derive an undirected graph by removing all non-mutual edges and the isolated nodes. We call the undirected graph Twitter network in this paper. It contains $190,520$ nodes and $1,001,528$ undirected edges. 2) DBLP dataset~\cite{snap_dblp} which is a co-authorship network in computer science where two authors are connected if they publish at least one paper together. Each node represents an author and each edge indicates co-authorship of a paper. There are totally 317,080 nodes and 1,049,866 undirected edges in DBLP network. 

\begin{figure}[h]
  \includegraphics[width=3in]{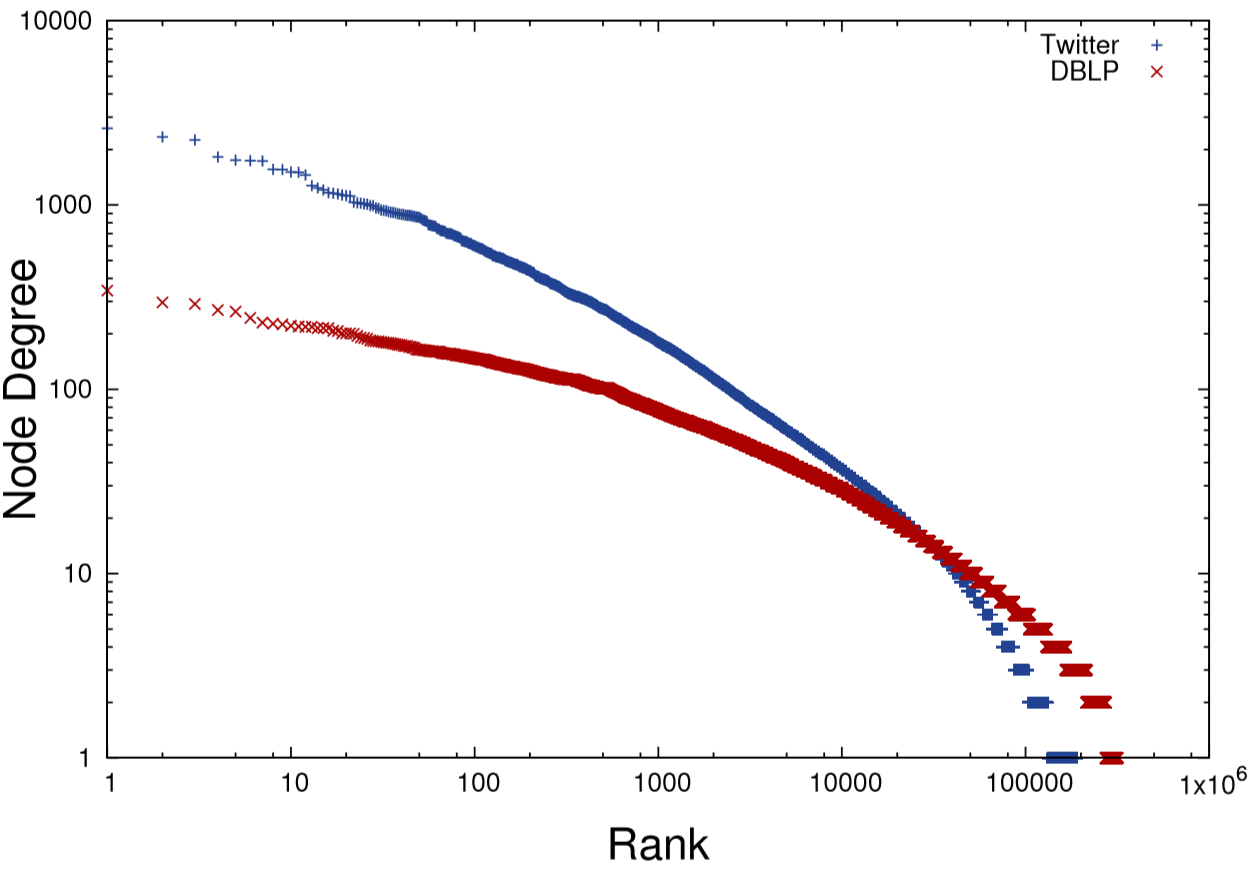}
  \caption{Node degree distribution of Twitter network and DBLP network}
  \label{fig:degree}
\end{figure}

\begin{figure}[h]
  \includegraphics[width=3in]{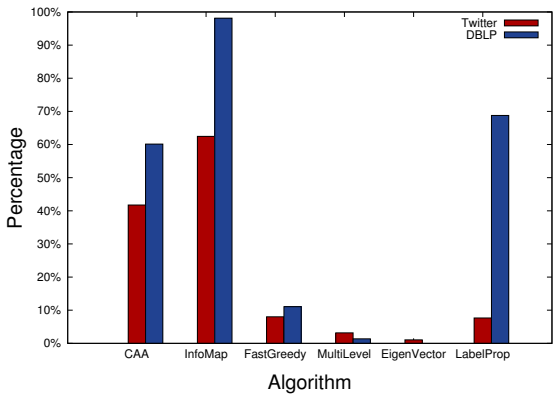}
  \caption{Desirable Community Coverage}
  \label{fig:coverage}
\end{figure}

To better understand the characteristic of these two networks, we plot the log-log graph for their node degree distribution in Fig.~\ref{fig:degree}. X-axis is the rank of a node, y-axis is the degree of the node, and a node with larger degree has lower rank. It is clear that node degrees in both networks follow power law distribution. Furthermore, these two networks are different in density and users in Twitter network have more neighbors than users in DBLP network in general. 

\begin{table*}[ht]
\centering
\caption{Summary of Algorithms}
\label{tab:algorithmInfo}\begin{tabular}{|c|c|c|c|} \hline
	Name & Overlap  & Time Complexity & Category \\ \hline
	CAA   &  Yes        & $Exponential, \, \, fast \, \, in \, \, practice \, \, for \, \, small \, \, cliques$            & Clique based \\ \hline
	Infomap & No      & $O(E)$     & Network coding and random walk  \\ \hline
	Multilevel  & No   & $O(N\log{}N)$       & Modularity maximization \\ \hline
	Eigenvector & No & $O(N(E+N))$  & Modularity maximization \\ \hline
	FastGreedy & No  & $O(N\log^{2}({N}))$ & Modularity maximization \\ \hline
	Label Propagation & No & $O(E)$ & Node labeling \\ \hline
\end{tabular}

\end{table*}

\subsection{Comparison Of Community Detection Algorithms}
As we mentioned in Section II, although a lot of research effort has been devoted to the comparison of different community detection algorithms and proposed different evaluation metrics, they ignore a very important factor, the size of the community. In this section, we first study the size distribution of the detected communities by different algorithms and calculate the percentage of users assigned to a desirable community of size between 4 to 150. Then we divide the community by their size into four groups of [4-50], [51-150], [151-500], and [501+] and compare the quality of communities in each group with the following criteria: extended modularity, triangle participation ratio (TPR), conductance, internal density, and transitivity. To make our results reproducible, we adopt the implementation of community detection algorithms in open-access python package igraph and NetworkX respectively. More specifically, we run Infomap, Multilevel, FastGreedy, Eigenvector, Label Propagation, Edge Betweenness, WalkTrap, and SpinGlass algorithms in igraph, Clique Percolation Method (CPM)~\cite{Palla2005} algorithm in NetworkX, and CAA algorithm on both Twitter network and DBLP network. In our experiment, we set the growing threshold of CAA to 0.7 and its overlapping threshold to 0. The communities grow from cliques of size $\ge 3$.  However, only Infomap, Multilevel, FastGreedy, Eigenvector, Label Propagation, and CAA can handle the scale of the data. Other algorithms hang due to their time complexity and the limitation of the hardware configuration of our experiment environment. Table~\ref{tab:algorithmInfo} summarizes the six algorithms that finish running. These algorithms can be classified into four categories based on their design ideas as follows: clique based (CAA), network coding and random walk (Infomap), modularity maximization (Multilevel, FastGreedy, Eigenvector), and node labeling (Label Propagation). In Table~\ref{tab:algorithmInfo}, overlap indicates whether an algorithm can produce overlapping communities which is more desirable for social media since a user in social media usually belongs to multiple communities. Overlapping community detection is not as extensively studied as disjoint algorithms, thus we are unable to find open source implementation of overlapping algorithms to compare besides CPM in NetworkX. However, we are unable to run CPM on our datasets due to the scale of the data. 

\begin{table}[htbp]
\centering
\caption{Total number of detected communities and the size of the largest community in Twitter network}
\label{tab:totalcommaz}
\resizebox{\columnwidth}{!}{%
\begin{tabular}{|c|c|c|} \hline
   Algorithm & Number of Communities & Largest Community Size / Ratio \\ \hline
    CAA & 12,312 & 680 / 0.35\% \\ \hline
    Infomap & 18,537 & 13,126 / 6.89\% \\ \hline
    Multilevel & 7,409 & 34,955 / 18.35\% \\ \hline
    Eigenvector & 5,834 & 136,403 / 71.6\% \\ \hline
    FastGreedy & 9,350 & 51,781 / 27.18\% \\ \hline
    Label Propagation & 7,648 & 149,022 / 78.22\% \\ \hline
\end{tabular}}
\end{table}

\begin{table}[htbp]
\centering
\caption{Total number of detected communities and the size
of the largest community in DBLP network}
\label{tab:totalcommdblp}
\resizebox{\columnwidth}{!}{%
\begin{tabular}{|c|c|c|} \hline
	Algorithm 	& Number Of Communities & Largest Community Size / Ratio  \\ \hline
	CAA     	& 28,213				& 270 / 0.085\%   \\ \hline
	Infomap 	& 16,999				& 587 / 0.185\%   \\ \hline
	Multilevel  & 565 					& 30,427 / 9.6\%   \\ \hline
	Eigenvector & 2  					& 315,569 / 99.5\%    \\ \hline
	FastGreedy  & 3,206					& 54,787 / 17.28\% \\ \hline
	Label Propagation & 21,156  & 84,363 / 26.6\% \\ \hline
\end{tabular}}
\end{table} 

\begin{table}[htbp]
\caption{Community Size Distribution in Twitter network}
\label{tab:communitesPerSizeAZ}
\resizebox{\columnwidth}{!}{%
\begin{tabular}{|c|c|c|c|c|c|c|} \hline
	Size Range & CAA    & Infomap & Multilevel & Eigenvector & FastGreedy & Label Propagation \\ \hline
	1 - 3		& 241 	   & 8,811       & 6,612      & 5,468          &  7,416     & 6,039 \\ \hline
	4 - 50        & 11,359 & 9,394   & 754        & 364         & 1,889      & 1,546    \\ \hline
	51 - 150   & 641 	& 223     & 11         & 0           & 20          & 36 \\ \hline
	151 - 500  & 69 	& 91      & 8          & 0           & 10          & 22 \\ \hline
	501+       & 2 	    & 18      & 24         & 2           & 15          &  5 \\ \hline
\end{tabular}}
\end{table}

\begin{table}[htbp]
\caption{Community Size Distribution in DBLP network}
\label{tab:communitesPerSizeDBLP}
\resizebox{\columnwidth}{!}{%
\begin{tabular}{|c|c|c|c|c|c|c|} \hline
	Size Range & CAA    & Infomap & Multilevel & Eigenvector & FastGreedy  & Label Propagation \\ \hline
	1 - 3	 	& 578      & 757       &  0             & 0                 & 258            & 3,065 \\ \hline
	4 - 50     & 27,176 & 15,475  & 415        & 0           & 2,782      & 17,737  \\ \hline
	51 - 150   & 454 	& 751     & 5          & 0           & 85          & 329 \\ \hline
	151 - 500  & 5 	    & 14      & 18         & 0           & 35          & 24 \\ \hline
	501+       & 0	    & 2       & 127        & 2           & 46          & 1 \\ \hline
\end{tabular}}
\end{table}

\subsubsection{Community Size Distribution} 

Table~\ref{tab:totalcommaz} and~\ref{tab:totalcommdblp} summarizes the number of communities and the size of the largest community revealed by each algorithm. We also include the percentage of users that are assigned to the largest community. As can be seen, modularity maximization algorithms (Multilevel, FastGreedy, Eigenvector) and Label Propagation all produce extremely large communities for both networks. For example, the largest community Eigenvector produces in Twitter network has size of $136,403$, that is, over $70\%$ of all users are grouped into one large community. In DBLP network, Eigenvector algorithm produces largest community of size $315,569$, that is $99.5\%$ of all users are grouped into one large community. There lack of strong connections among community users in such large communities and we can hardly put communities of such large size to practical use.  

Table~\ref{tab:communitesPerSizeAZ} and~\ref{tab:communitesPerSizeDBLP} provide more detailed breakdown of the number of communities in different size ranges. As can be seen, CAA and Infomap produce more communities in desirable size of 4-150 than others. Another interesting observation is that Label Propagation algorithm performs significantly better on DBLP network than on Twitter network. Eigenvector performs poorly on both networks, especially on DBLP network since all it discovers are two large communities of size greater than 501. As we can see in Figure 3-7, Eigenvector does not have any score in range [4-50], [51-150] and [151-500] for DBLP network since it does not generate communities in those size ranges.  From the community size distribution point of view, we would recommend CAA and Infomap which can produce decent number of communities with desirable sizes.   

\subsubsection{Community Coverage} 
In this section, we define a new metric called \emph{desirable community coverage} to measure the number of users assigned to desirable communities, that is, communities of size 4 to 150. Generally speaking, the higher the desirable community coverage, the better the algorithm. Note that even Infomap, Multilevel, Eigenvector, and FastGreedy all assign every single node in the graph to a community, they do not provide $100\%$ coverage in our definition of desirable community coverage. Fig.~\ref{fig:coverage} shows the performance of each algorithm with regard to desirable community coverage while taking the community size into account.  


It is clear that Infomap performs the best in terms of desirable community coverage since it assigns $62\%$ of all users in the Twitter network and $98\%$ of all users in DBLP network into desirable communities. CAA also performs well in both networks when compared to other methods. It is worth noting that in this experiment, CAA starts from non-overlapping cliques by setting the overlapping threshold to 0. In fact, CAA can achieve higher coverage if we allow overlapping cliques to start with. Label Propagation performs inconsistently in two networks and FastGreedy, Multilevel, and Eigenvector failed to assign the majority of users in the network to meaningful communities.

\subsubsection{Extended Modularity} 
Extended modularity has been defined as a popular metric to measure the goodness of overlapping communities in~\cite{Huawei2009}. We give the definition of this metric in Equation~\eqref{eq:extendedModularity}, where $C_i$ is the $ith$ community, $O_{v}$ is the number of communities the vertex $v$ belongs to, similarly $O_{w}$ is the number of communities vertex $w$ belongs to. $A$ is the adjacency matrix, that is, $A_{vw} = 1$ means there exists an edge between vertex $v$ and vertex $w$, otherwise it is 0. $\frac{k_v k_w}{2m}$ describes the expected number of edges between vertex $v$ and vertex $w$. $k_v$ and $k_w$ are the node degree of $v$ and $w$ in the whole topology respectively. $m$ is the total number of edges in the whole topology. The range of extended modularity is $[-1,1]$ and the higher the value, the better the community in terms of modularity. The intuition behind extended modularity is that communities should have more internal connectivity than random graph of the same degree sequence. Extended modularity is a variation to modularity, which is the most widely used metric to measure the goodness of disjointed community. If there is no overlap between the communities, $\frac{1}{O_v O_w}$ is 1 and Equation \eqref{eq:extendedModularity} becomes the traditional modularity as defined in~\cite{Clauset2004}. In general, overlapping communities have lower modularity than disjoint communities since overlapping communities have many connections outside the community.
\begin{displaymath}
  \displaystyle EQ = \frac{1}{2m} \sum_{i} \sum_{v \in C_i, w \in C_i }  \frac{1}{O_v O_w}\Bigl[ A_{vw} - \frac{k_v k_w}{2m} \Bigl] 
  \tag{1}
  \label{eq:extendedModularity}
\end{displaymath}

\begin{figure*}[h!]
\centering
\begin{subfigure}[b]{0.44\textwidth}
\includegraphics[width=\textwidth]{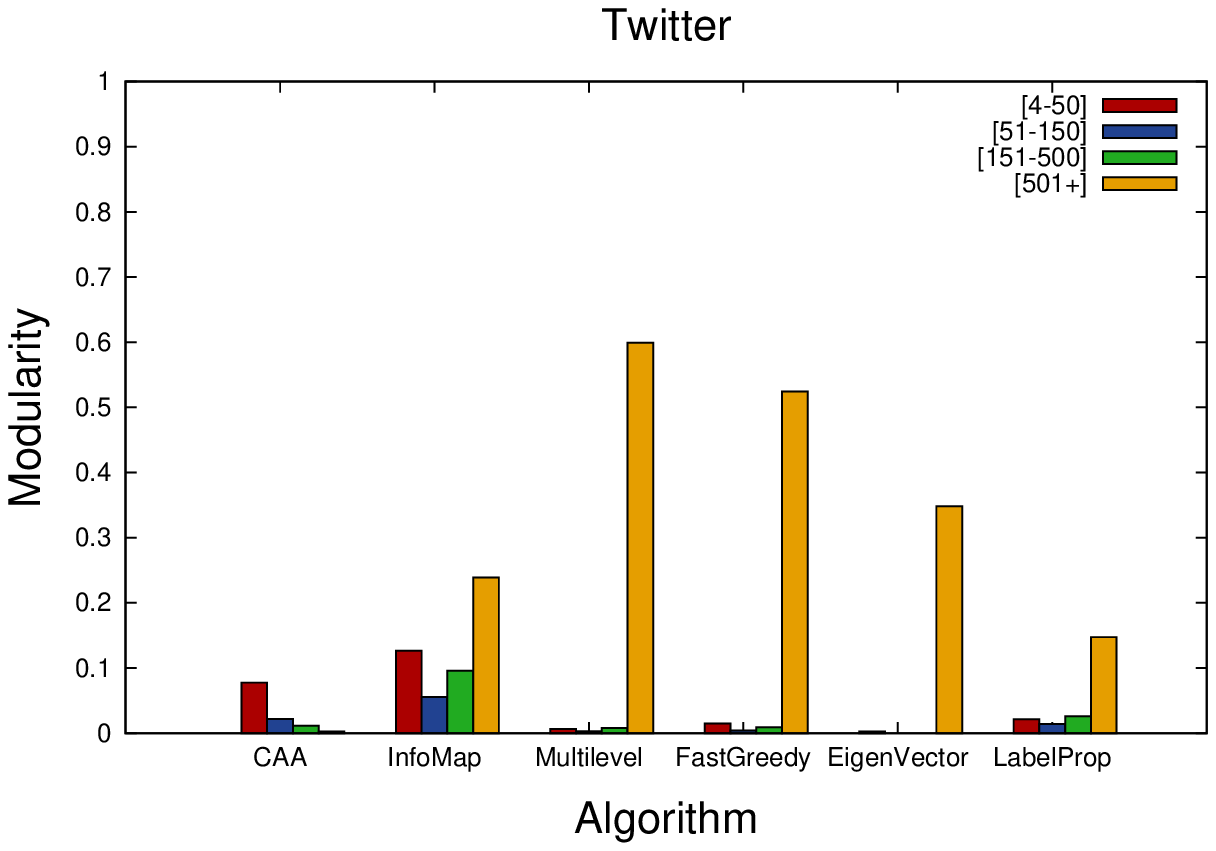}
\end{subfigure}
\begin{subfigure}[b]{0.44\textwidth}
\includegraphics[width=\textwidth]{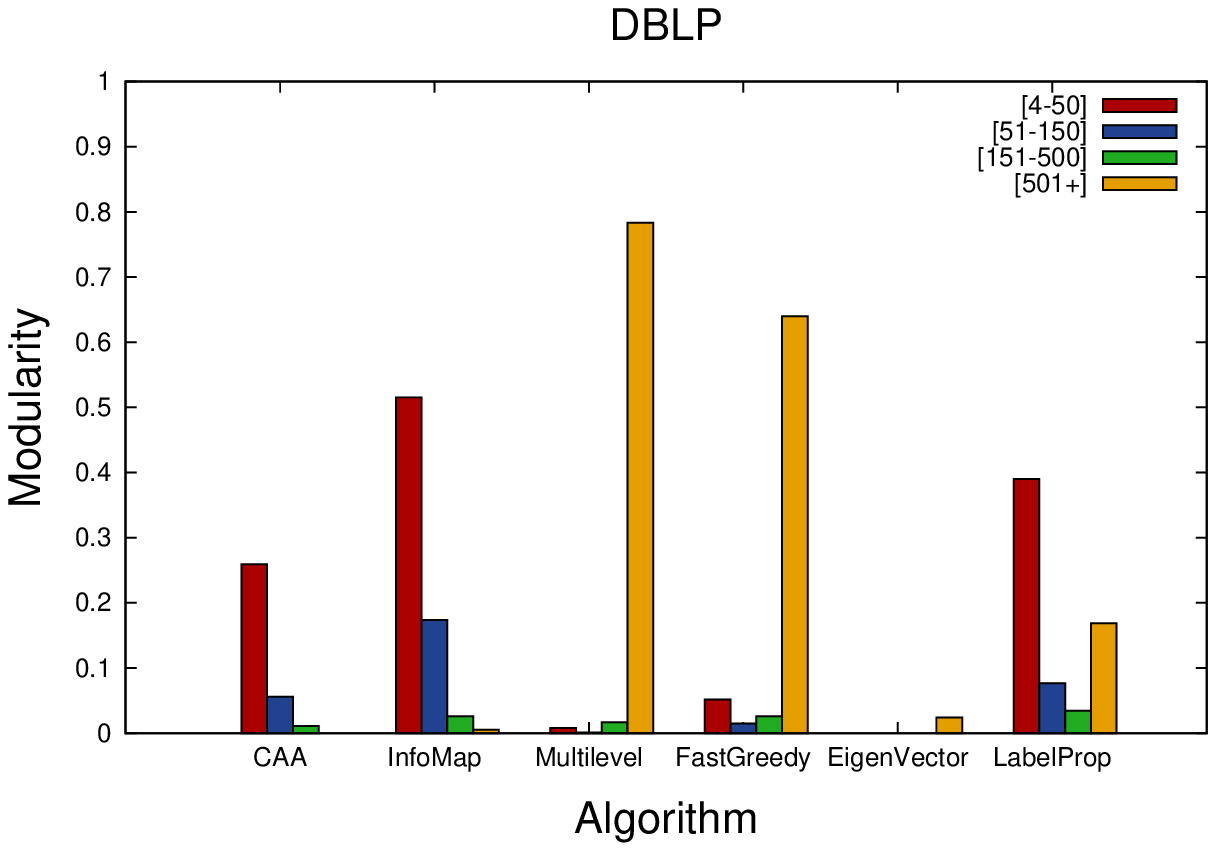}
\end{subfigure}
\caption{Extended Modularity}
\label{fig:extendedModularity}
\end{figure*}

\begin{figure*}[h!]
\centering
\begin{subfigure}[b]{0.44\textwidth}
\includegraphics[width=\textwidth]{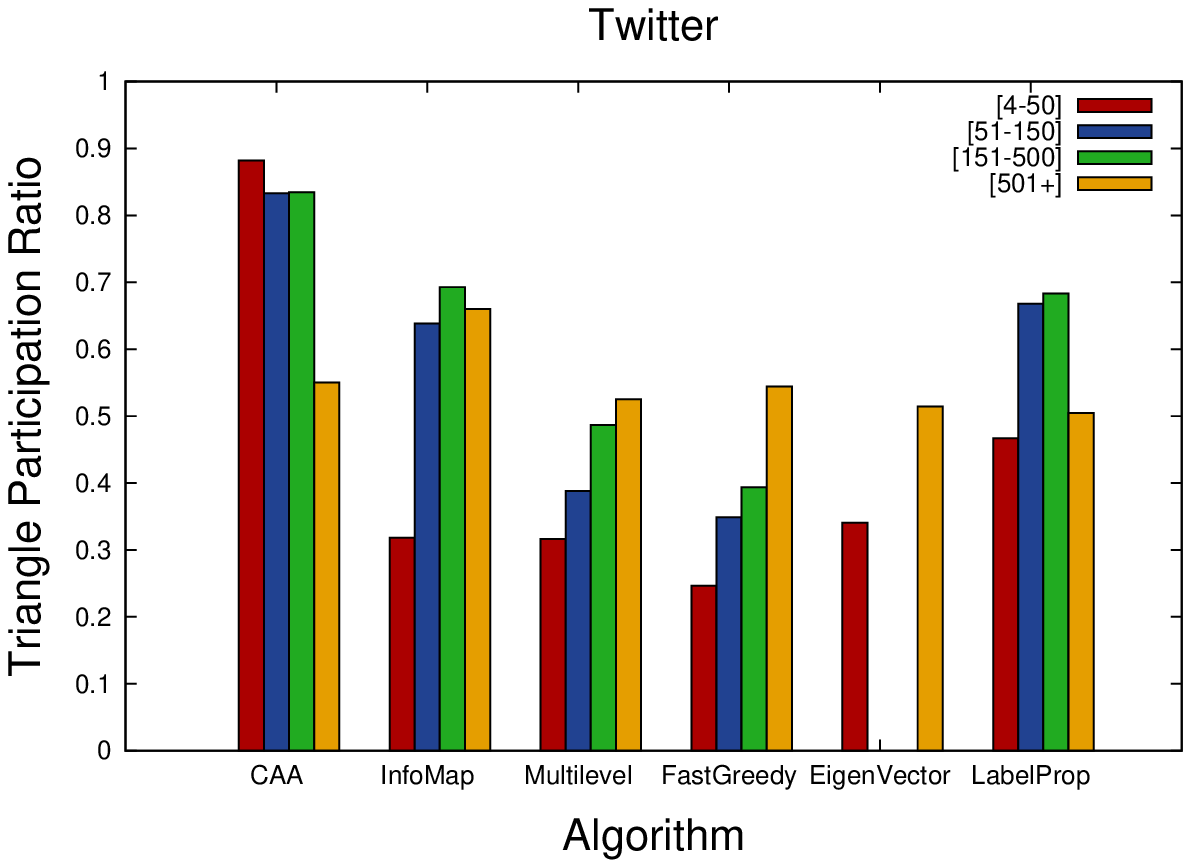}
\end{subfigure}
\begin{subfigure}[b]{0.44\textwidth}
\includegraphics[width=\textwidth]{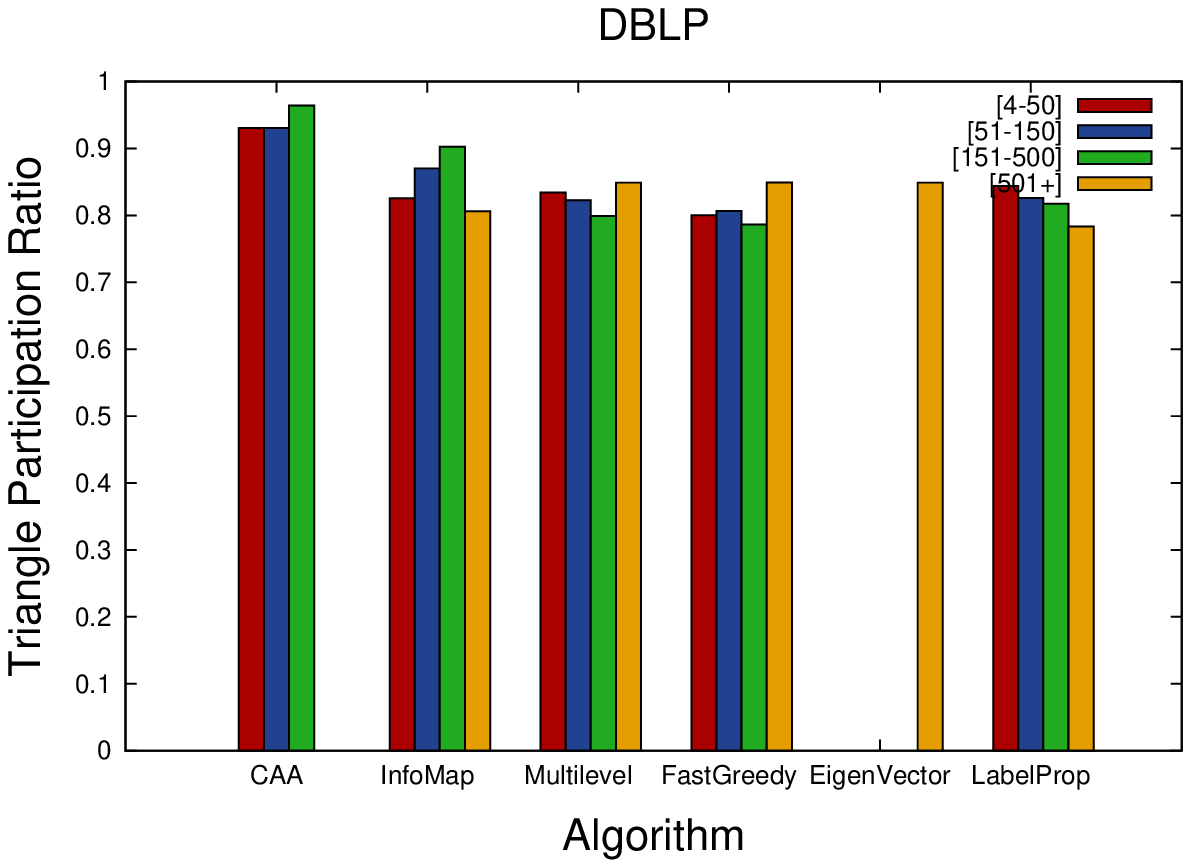}
\end{subfigure}
\caption{Triangle Participation Ratio}
\label{fig:tpr}
\end{figure*}

As can be seen from Equation \eqref{eq:extendedModularity}, each community contributes a value towards the modularity score. To measure to what degree communities of different sizes impact the modularity, we divide the sum in Equation \eqref{eq:extendedModularity} into four parts, that is, the sum over all communities of size [4-50], [51-150], [151-500], and [501+] and present the partial modularity score contributed by each community group in Figure~\ref{fig:extendedModularity}. Infomap outperforms others because unlike Multilevel, FastGreedy, and Eigenvector, whose modularity scores are mostly decided by the large communities of size [501+], Infomap achieves high score that is contributed by all four groups of it's communities. In fact, its modularity score is very close to the highest score achieved by Multilevel. The observation that larger communities achieve higher modularity score is consistent with the resolution limit of modularity as indicated in~\cite{Fortunato2007}. This discovery also suggests we should maximize modularity while control the community size at the same time. It is also worth noting that in general overlapping communities have lower modularity than non-overlapping ones thus the low modularity of CAA is expected behavior. 

\begin{figure*}[h]
\centering
\begin{subfigure}[b]{0.44\textwidth}
\includegraphics[width=\textwidth]{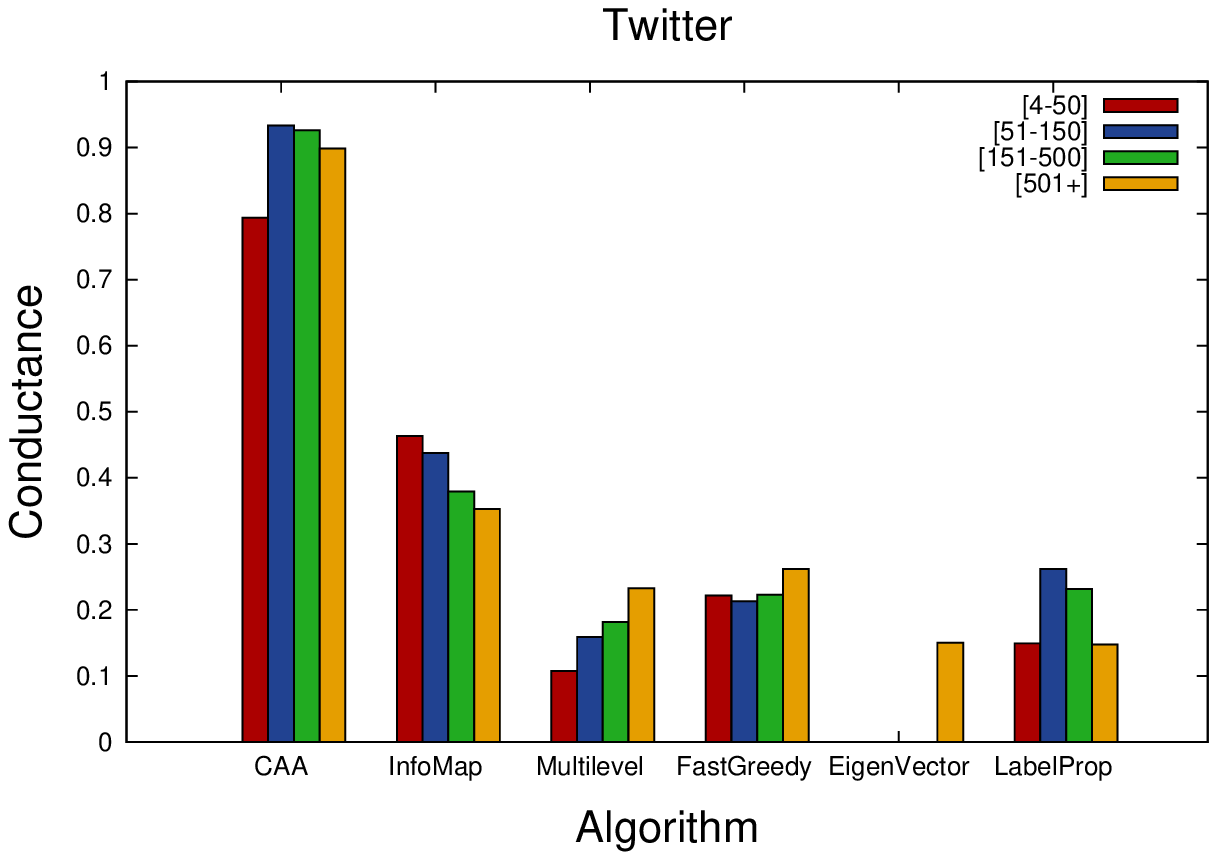}
\end{subfigure}
\begin{subfigure}[b]{0.44\textwidth}
\includegraphics[width=\textwidth]{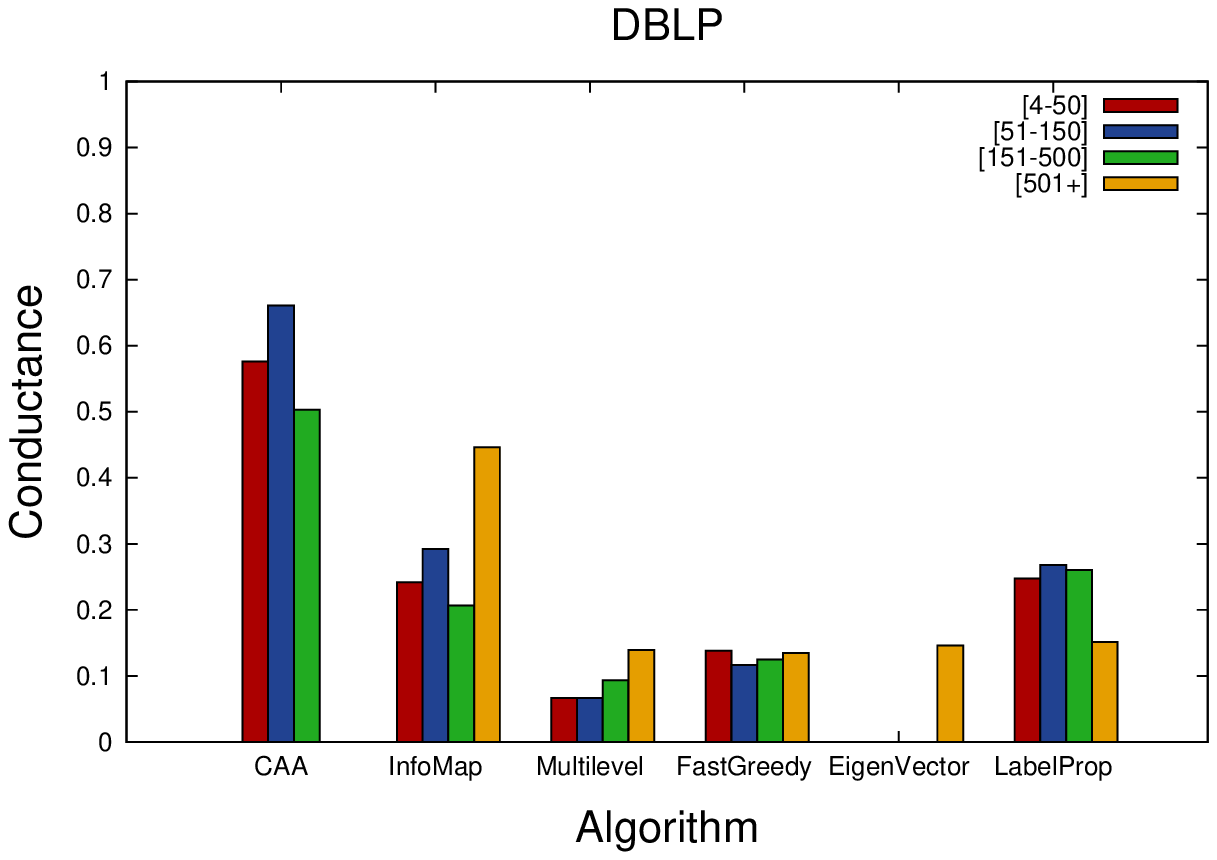}
\end{subfigure}
\caption{Conductance}
\label{fig:cond}
\end{figure*}

\begin{figure*}[h]
\centering
\begin{subfigure}[b]{0.44\textwidth}
\includegraphics[width=\textwidth]{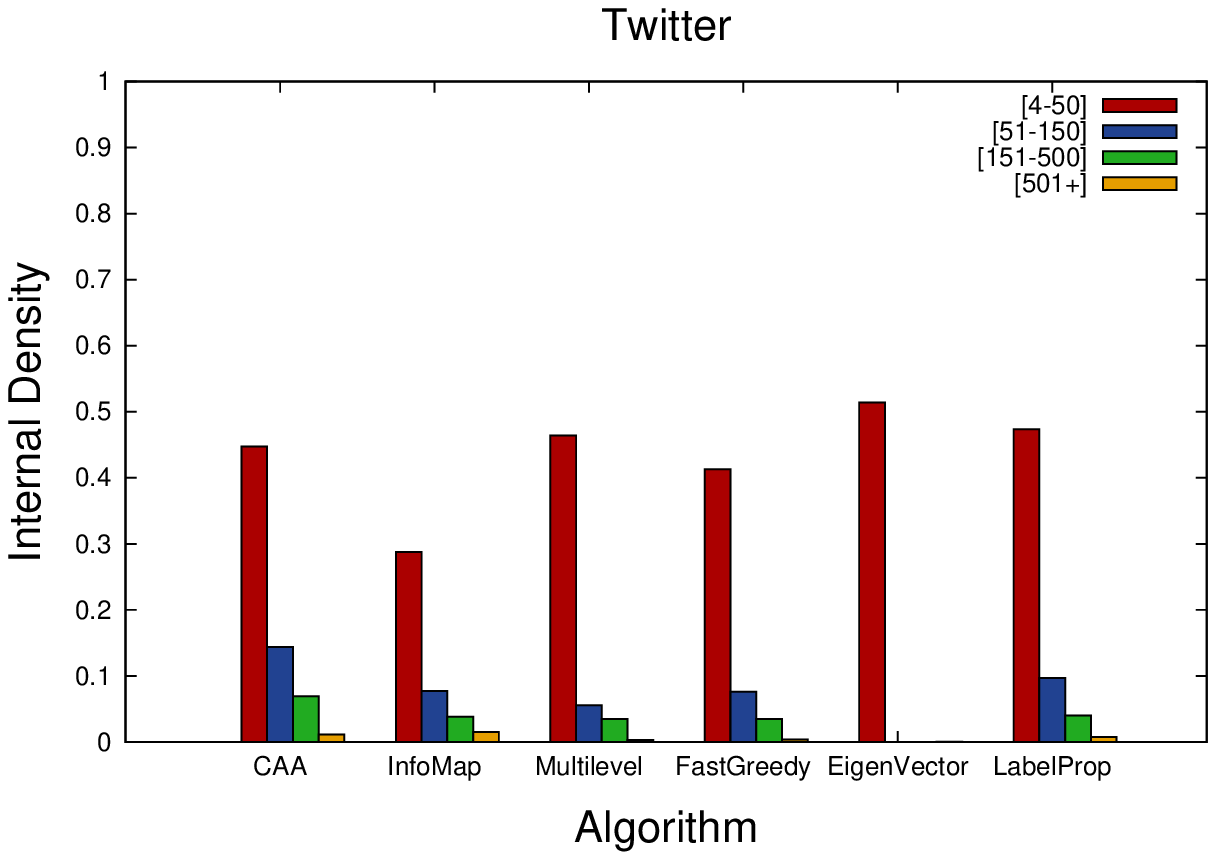}
\end{subfigure}
\begin{subfigure}[b]{0.44\textwidth}
\includegraphics[width=\textwidth]{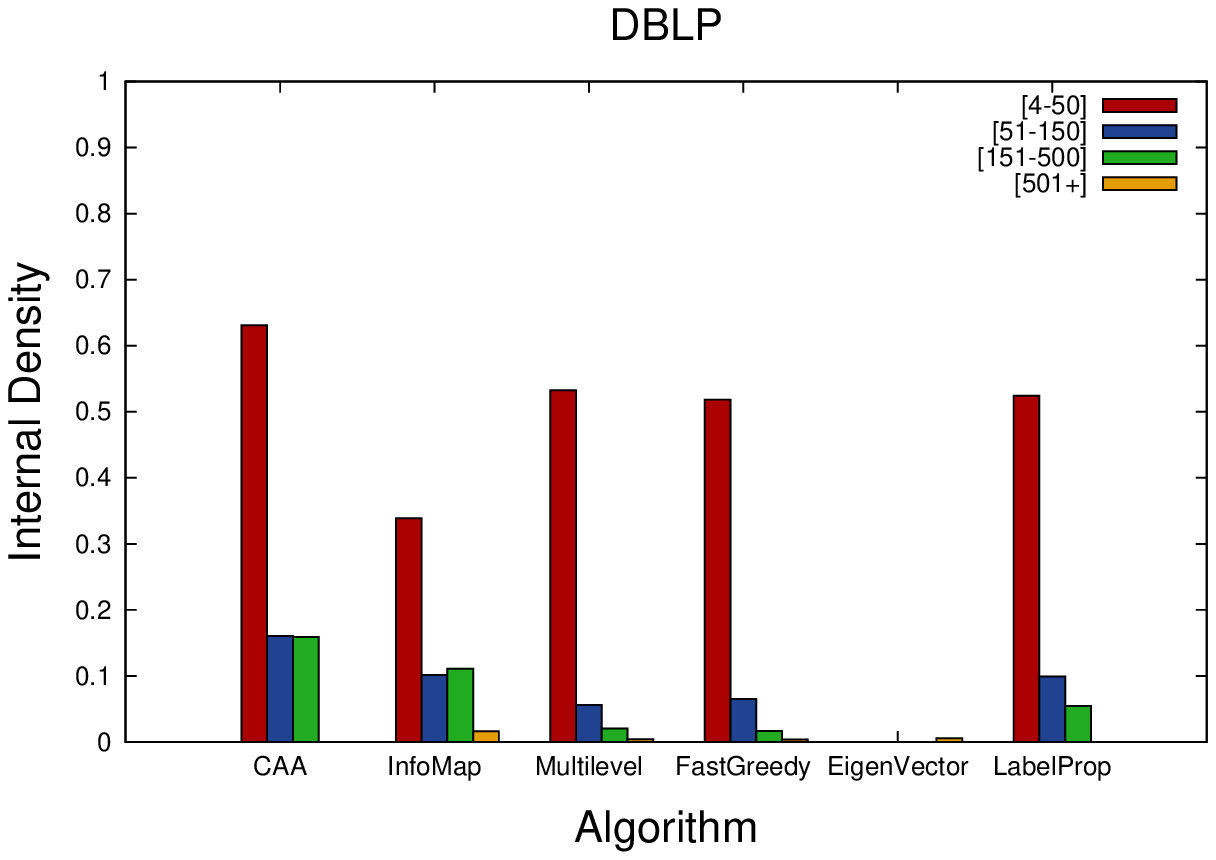}
\end{subfigure}
\caption{Internal Density}
\label{fig:internal}
\end{figure*}

\subsubsection{Triangle Participation Ratio}
TPR was proposed in~\cite{Yang2012} as a metric for community evaluation, where it is defined as the number of nodes in a community that form a triad, divided by the total number of nodes in the community.  ~\cite{Yang2012} found that the communities discovered by algorithms to optimize community modularity do not align with the ground truth communities, that is, the known membership in real life, while TPR is a good metric when looking for ground-truth communities.  As indicated by Fig.~\ref{fig:tpr}, CAA achieves significantly higher TPR than others on both networks, with close to 0.9 TPR score on Twitter network and above 0.9 TPR score on DBLP network. This is expected since CAA starts from a clique with TPR as 1 and grows the clique with new nodes that are highly connected to the growing community.  We can also see than all algorithms show satisfactory performance on DBLP network in terms of TPR. 

\begin{figure*}[h]
\centering
\begin{subfigure}[b]{0.44\textwidth}
\includegraphics[width=\textwidth]{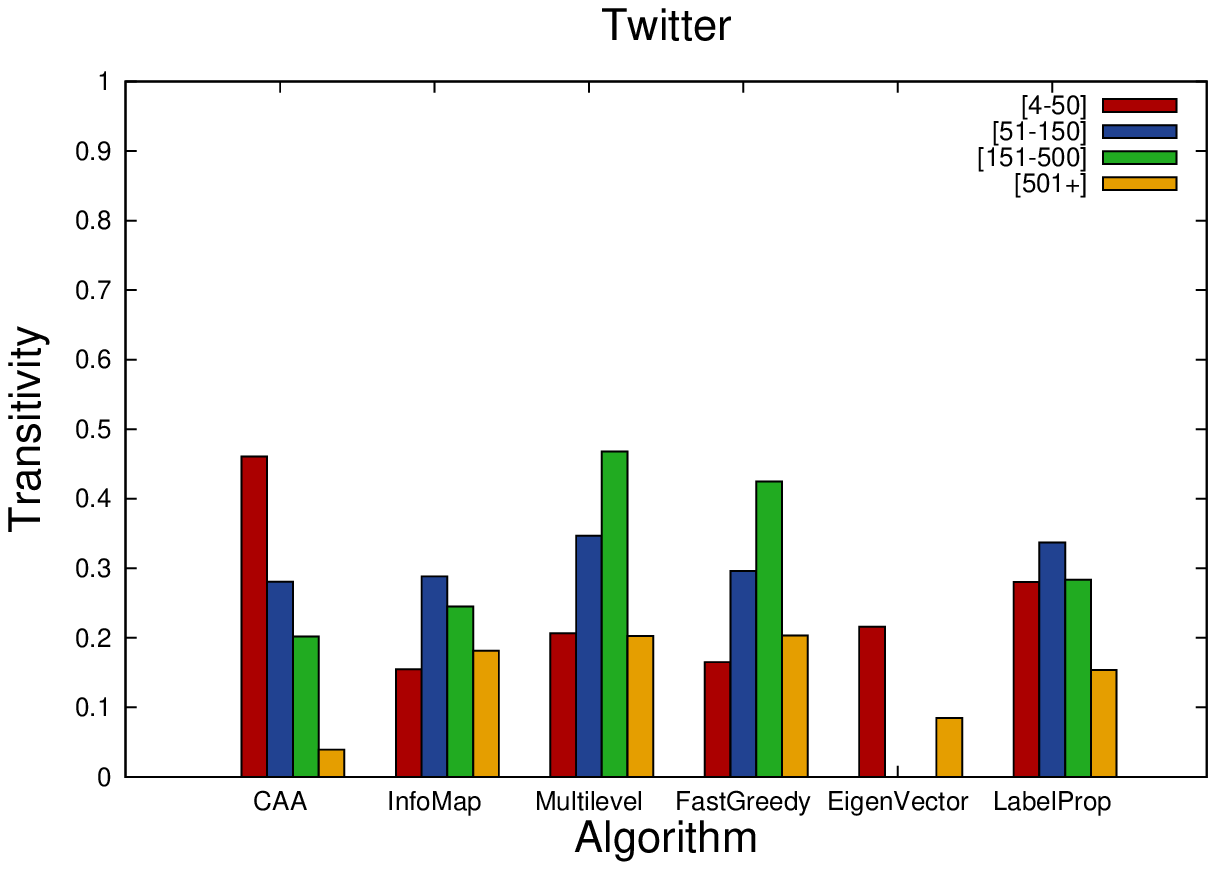}
\end{subfigure}
\begin{subfigure}[b]{0.44\textwidth}
\includegraphics[width=\textwidth]{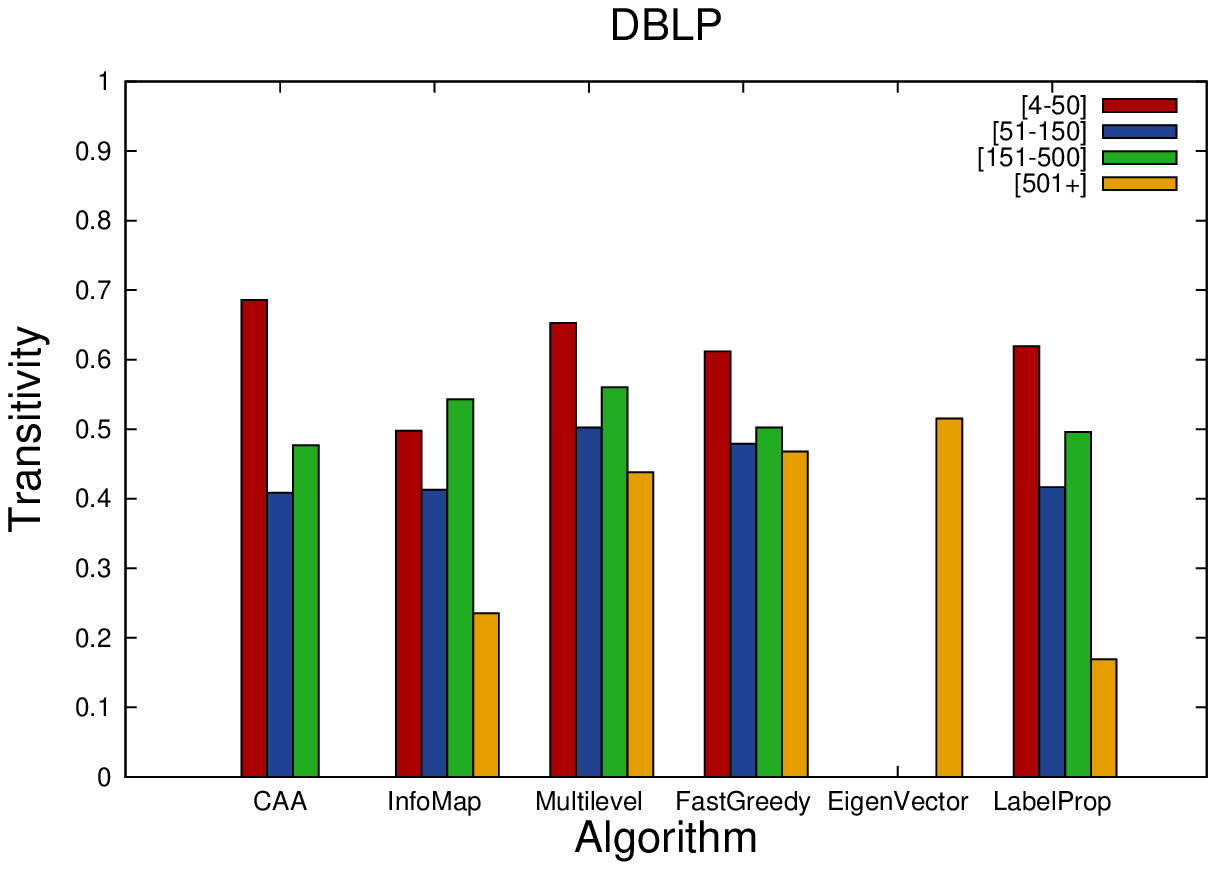}
\end{subfigure}
\caption{Transitivity}
\label{fig:transitivity}
\end{figure*}

\subsubsection{Conductance}
Conductance is a metric that takes into consideration both external and internal connections of a community~\cite{Jure2010}. It is defined as the ratio of the number of edges on the boundary of the community over the sum of degrees of nodes in the community. Lower conductance score indicates better community. We calculate the average conductance of communities in different size groups in Fig.~\ref{fig:cond}. We find that Multilevel achieves the lowest average conductance score and all three modularity maximization algorithms perform well. CAA performs poorly in conductance as expected since it produces overlapping communities. 

\subsubsection{Internal Density}
Internal density~\cite{Jure2010} is defined in Equation \eqref{eq:internalDensity}, where $m_{S}$ is the number of edges in the subset $S$ divided by the total possible edges between all nodes $n_{S}(n_{S}-1)/2$. Internal density is a measurement of the internal structure within the community.

\begin{displaymath}
  \displaystyle f(S) = \frac{m_{S}}{n_{S}(n_{S}-1)/2}
  \tag{3}
  \label{eq:internalDensity}
\end{displaymath}

\begin{figure*}[ht!]
\centering
\begin{subfigure}[b]{0.44\textwidth}
\includegraphics[width=\textwidth]{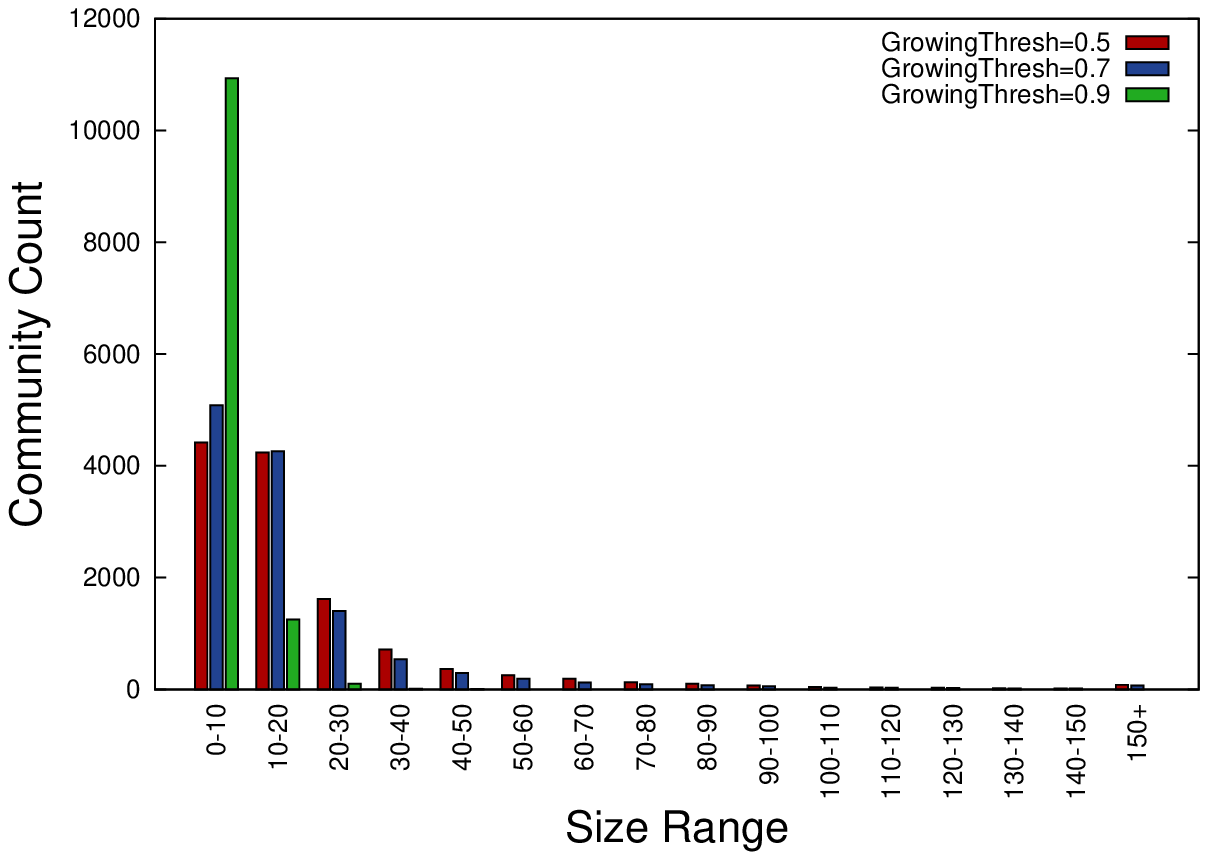}
\caption{The Effect of Growing Threshold}
\label{fig:growingThreshold}
\end{subfigure}
\begin{subfigure}[b]{0.44\textwidth}
\includegraphics[width=\textwidth]{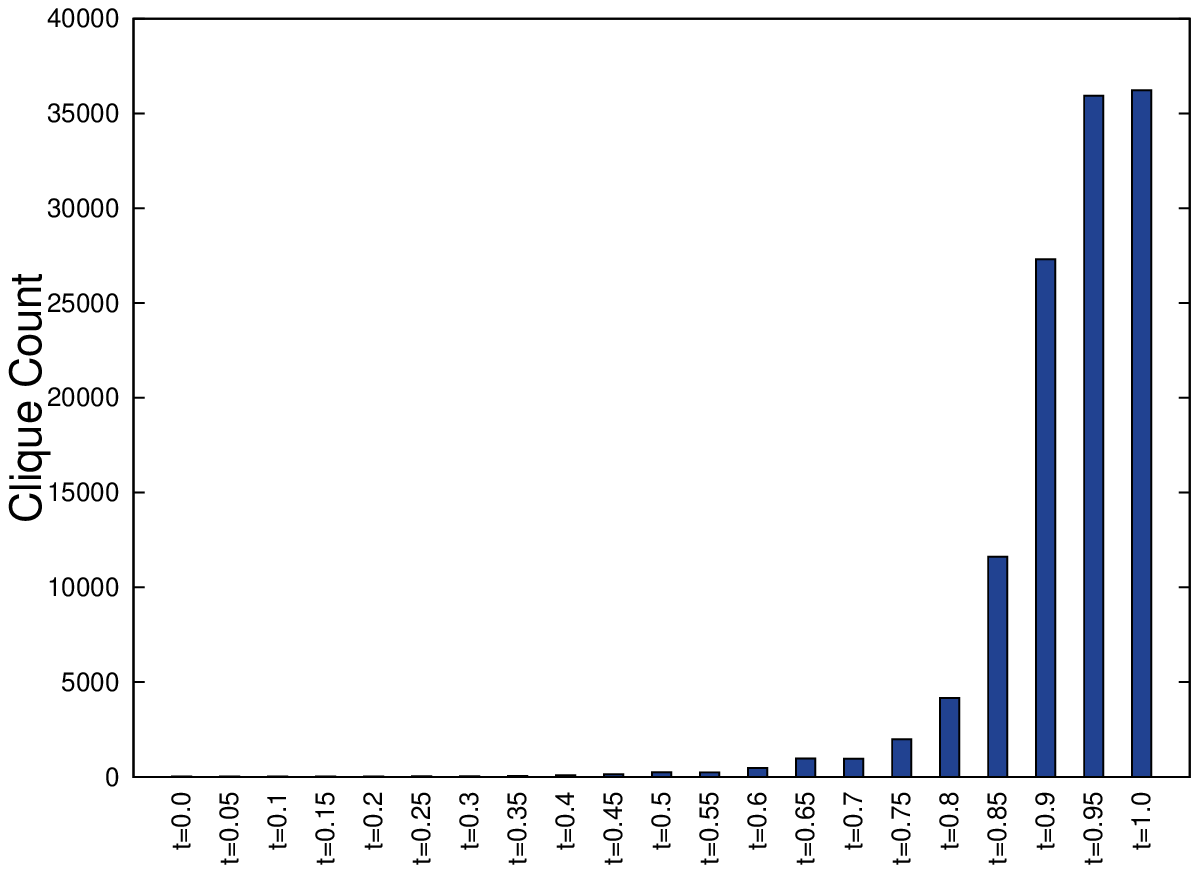}
\caption{The Effect of Overlapping Threshold}
\label{fig:overlappingThreshold}
\end{subfigure}
\caption{CAA Parameter Analysis}
\end{figure*}

Fig.~\ref{fig:internal} gives average internal density for communities in each size group. We find that smaller community size typically indicates higher internal density value. This is not surprising as smaller community sizes tend to form cliques. All algorithms show similar performance and CAA consistently performs slightly better across all size groups on DBLP network. 

\subsubsection{Transitivity}
Transitivity~\cite{EdgeBetweeness2002}  measures ties between individuals and is defined in Equation \eqref{eq:transitivityEq}.  This value indicates the probability that two friends of a single person are also friends. A clique has transitivity of 1.0.  Higher transitivity value is better since it indicates that the community is more tightly connected to one another and has a higher probability that users within the community know each other.
\begin{equation}
  \displaystyle \frac{\text{3 $\times$ number of triangles}}{\text{number of triads}}
  \tag{4}
  \label{eq:transitivityEq}
\end{equation}

The result of the average Transitivity score is presented in Fig.~\ref{fig:transitivity}. We notice that CAA performs the best in range [4-50] for both networks. All algorithms perform similarly in range [51-150]. 

In summary, the experimental results suggest that Infomap and CAA outperform others in terms of both community size and coverage by consistently producing desirable communities covering a large portion of users in both Twitter and DBLP networks. Nevertheless, both modularity maximization algorithms including Multilevel, FastGreedy, and Eigenvector and node labeling algorithm Label Propagation tend to output extremely large communities of size over 30,000. In terms of modularity, InfoMap performs the best in general. The identified community of different sizes all contribute to the modularity score and its total modularity is comparable to modularity maximization algorithms. For conductance, modularity maximization algorithms perform better than other algorithm indicating their identified communities are better separated from the external compared to other algorithms. With regard to triangle participation ratio, CAA outperforms others reaching above $90\%$ in DBLP network and close to $90\%$ in Twitter network. All algorithms perform similarly in terms of internal connectivity metrics, transitivity and internal density, with CAA performs slightly better for desirable community sizes. 

\subsection{Growing Threshold and Overlapping Threshold} \label{ssec:growOverlap}

In this section, we measure the impact of growing threshold and overlapping threshold on community size and the number of communities in order to give suggestions on the parameter selection for CAA algorithm. 

First we look into the effect of growing threshold. We find all cliques of size 3 and larger in Twitter network and set overlapping threshold to 0 to find all non-overlapping cliques. We set the growing threshold to 0.5, 0.7, and 0.9 indicating a neighboring node can only be added to the community if it is connected to at least 50\%, 70\%, or 90\% of nodes in the community. The result is plotted in Fig.~\ref{fig:growingThreshold}. x-axis is the community size range and y-axis is the number of communities whose size fall in the range. It can be seen that the cliques do not grow much with growing threshold $0.9$ since most communities are of size between 3 and 9. With growing threshold set to $0.7$, the number of communities with size in range [3-9] drops significantly, from around 11,000 to 5,000 and there are more communities in the range of [10-150]. Therefore, we recommend to set the growing threshold to $0.7$. Another interesting observation is that there is no significant difference in the distribution of community size for growing threshold $0.5$ and $0.7$. 

Next we investigate the effect of overlapping threshold. We choose all cliques of size over 15 and increase the overlapping threshold from 0 to 1. Intuitively, by increasing the overlapping threshold, less cliques are filtered, therefore the number of communities increases. As can been seen in Fig.~\ref{fig:overlappingThreshold}, where x-axis is the overlapping threshold value, y-axis is the number of cliques, the number of cliques increases significantly for overlapping threshold $\geq 0.8$. In general, we suggest to choose overlapping threshold less than 0.6 to avoid having heavily overlapping communities.  

\section{Related Work}
A lot of effort has been devoted to the area of community detection along with ways to determine the quality of the identified communities. Existing community detection algorithms can be categorized into disjoint algorithms and overlapping algorithms, based on whether the identified communities have overlap or not. Infomap~\cite{Infomap2008} stands out as the most popular and widely used disjoint algorithm. It is based on random walks on networks combined with coding theory with the intent of understanding how information flows within a network. Multilevel~\cite{MultilevelBlonde2008} is a heuristic based algorithm based on modularity optimization. Multilevel first assigns every node to a separate community, then selects a node and checks the neighboring nodes attempting to group the neighboring node with the selected node into a community if the grouping results in an increase in modularity. Newman's Leading Eigenvector~\cite{EigenVector2006} works by moving the maximization process to the eigenspectrum to maximize modularity by using a matrix known as the modularity matrix. Fast Greedy~\cite{FastGreedy2004} is based upon modularity as well. It uses a greedy approach to optimize modularity. Label propagation~\cite{LabelPropgatoin2007} algorithm works by assigning a unique label to each node in the graph. The nodes are then listed in a random sequential order which the algorithm follows to diffuse the labels through the network. This causes nearby neighboring nodes to adopt the same label from the nodes nearest to them. This process causes community like sets of nodes to quickly converge to a final label that uniquely identifies the group. 

In the category of overlapping algorithm, Clique Percolation Method \cite{Palla2005} merges two cliques into a community if they overlap more than a threshold. Another relevant paper \cite{Jierui2012} discusses overlapping community detection algorithms along with various quality measurements. 

One comprehensive survey of recent advances~\cite{Santo2010} discusses a wide range of existing algorithms including traditional methods, modularity based methods, spectral algorithms, dynamic algorithms, and more.  Community detection often tries to optimize various metrics such as modularity as described by Girvan and Newman \cite{Newman2004} or conductance. The work in \cite{Jure2010} discusses many of the various objective functions currently in use and how they perform. Similarly, \cite{Yang2012} points out that it is important for the community detection algorithm to extract functional communities based on ground truth, where functional ground-truth community is described as a community in which an overall theme exists. Another recent paper~\cite{YangNature2016} conducts a comparative analysis of community detection algorithms on LFR benchmark in order to provide unbiased guidelines to choose algorithms for a given network. They propose a method to quantify the accuracy of the algorithm with mixing parameter. In addition, \cite{WangVLDB2015} provides a procedure-oriented framework for benchmarking and compare different algorithms. The authors also fit and implement 10 algorithms in their framework and compare and rate the algorithms. 

Our work is different from existing comparative research in that we propose a new angle of comparing the structural metrics for community detection algorithms by considering the size of the identified communities. Our work is the first to investigate previously overlooked community size in a systematic way. 

\section{Conclusion}
In this paper, we carry out a comparative analysis on community detection algorithms by introducing community size as a new orthogonal dimension to existing structural metrics. In order to make sense of communities of different sizes, we introduce the Dunbar's number in sociology into the community detection in social media. We find that both Infomap and CAA are able to discover communities of desirable sizes. We also present and compare the performance of each evaluated algorithm in all evaluated structural metrics. 

Our future work includes exploring greedy approach to improve the performance of CAA in terms of conductance and modularity, conducting comparison with more overlapping community detection algorithms. Furthermore, we plan to explore new metrics such as whether and in what degree a community interest exists in the identified community for community quality evaluation and also make sense of the community.  


\end{document}